# Title: A versatile coherent Ising computing platform

## Authors

Hai Wei[1†*], Chengjun Ai[1†], Putuo Guo[1†], Bingjie Jia[1†], Lixin Yuan[1†], Hanquan Song[1], Shaobo Chen[1], Chongyu Cao[1], Jie Wu[1], Chao Ju[1], Yin Ma[1,2], Jintao Fan[3], Minglie Hu[3], Chuan Wang[1,2*], Kai Wen[1*]

## Affiliations

[1] Beijing QBoson Quantum Technology Co., Ltd., Beijing, 100016, China

[2] School of Artificial Intelligence, Beijing Normal University, Beijing, 100875, China

3 Ultrafast Laser Laboratory, School of Precision Instruments and Opto-electronics Engineering, Tianjin University, Tianjin, 300072, China

[†] These authors contributed equally to this work

[*] Corresponding authors: H. Wei weih@boseq.com, C. Wang wangchuan@bnu.edu.cn, and K. Wen wenk@boseq.com

## Abstract

Coherent Ising Machines (CIMs) have emerged as a hybrid form of quantum computing devices designed to solve NP-complete problems, offering an exciting opportunity for discovering optimal solutions. Despite challenges such as susceptibility to noise-induced local minima, we achieved notable advantages in improving the computational accuracy and stability of CIMs. We conducted a successful experimental demonstration of CIM via femto-second laser pumping that integrates optimization strategies across optical and structural dimensions, resulting in significant performance enhancements. The results are particularly promising. An average success rate of 55% was achieved to identify optimal solutions within a Möbius Ladder graph comprising 100 vertices. Compared with other alternatives, the femto-second pulse results in significantly higher peak power, leading to more pronounced quantum effects and lower pump power in optical fiber based CIMs. In addition, we have maintained an impressive success rate for a continuous period of 8 hours, emphasizing the practical applicability of CIMs in real-world scenarios. Furthermore, our research extends to the application of these principles in practical applications such as molecular docking and credit scoring. The results presented substantiate the theoretical promise of CIMs, paving the way for their integration into large-scale practical applications.

**Teaser**

First demonstration of stable Coherent Ising Machines as a computing platform for versatile applications.

**MAIN TEXT**

**Introduction**

Combinatorial optimization problems are problems seeking the optimal set of parameter values that maximize or minimizes a predefined utility function. Applications of combinatorial optimization are pervasive in many real-world scenarios such as scheduling, routing and resource allocation [1,2]. As the solution space increases exponentially with the number of variables, finding the optimal solution is challenging for most combinatorial problems. Researchers traditionally rely on approximation algorithms that provide near-optimal solutions in polynomial time [1,3]. Recently, many alternative approaches have been proposed to address this challenge and other similar issuses using physical systems, including quantum annealers [4], neural networks [5,6], ploaritons[7], resistive memory devices[8], and photonics [9,10,11]. Among these, one promising method is Coherent Ising

Machine (CIM) [2]. CIM is a hybrid quantum computing device that simulates the Ising Model [3] using Degenerate Optical Parametric Oscillator (DOPO) pulses [2,3,12–15]. A CIM typically consists of a network of DOPOs as spins [12,16–18], which are generated through non-linear optical processes and coupled pairwise to encode the Hamiltonian of the target optimization problem [14]. By precisely adjusting the pump power, CIM will transit from the below-threshold state to the above-threshold state, through which the state with minimum loss would oscillate and the ground state of the encoded Hamiltonian could be identified accordingly[19]. It has been demonstrated in multiple studies [20,21] that CIMs can accelerate the computation of Ising problems by more than 2 orders of magnitude compared to classical computers for problems up to 100,000 spins [15,16,22].

However, CIM faces challenges from high sensitivity to cavity length variations and environmental noises (including temperature fluctuations and mechanical vibrations), which can lead to suboptimal computing results [17,23,24]. Though many theoretical studies on error-mitigated computations in CIM have been presented [18,24,25], experimental demonstrations remain scarce [26].

In this study, we investigated various ways to improve the solution quality for Ising problems solved by CIMs. A novel approach using femto-second (fs) pulse laser was developed to achieve high-efficiency non-linearity during DOPO preparation. In conjunction with our new cavity stabilization system, this approach allows for DOPO generation at a lower average pump power with more pronounced quantum effects, leading to improved solution quality. We have experimentally demonstrated that the average probability of achieving the optimal solution for 100-vertex Möbius Ladder Max-Cut problem can be increased to 55%, which represents the highest reported success rate among CIM or other Noisy Intermediate-Scale Quantum (NISQ) systems. Moreover, over 8-hour continuous operation can be achieved in our CIM system, on which a cloud platform is set up to host requests from users of different industries. Lastly, two example computing tasks are performed to illustrate the potential of our fs CIM system for practical applications across different areas.

**Results**

**The setup of the CIM**

The Hamiltonian of the Ising model without an external magnetic field can be expressed as:

$$H(\sigma) = -\sum_{ij} J_{ij} \sigma_i \sigma_j \quad (1 \leq i, j \leq N) \tag{1}$$

where $\sigma_i \in \{+1, -1\}$ denotes the value of the $i$-th Ising spin, $J_{ij}$ is the interaction strength between the $i$-th and $j$-th spins, and $N$ is the total number of spins. As shown in previous works of CIM [13], each DOPO pulse represents a spin, with its phase value ($0$ or $\pi$) corresponding to the spin state in an Ising model. During computation, the pump power is gradually lifted from below the DOPO threshold to above. When the pumping power is below the threshold, each spin is in a superposition state, where the phase is in a superposition $0$ and $\pi$. When the pump power goes above the threshold, each spin would collapse into a determined state of $0$ or $\pi$. The pulses are optically interfered to realize couplings between different spins as defined by $\{J_{ij}\}$. According to the minimum gain principle [27], after transition to above the threshold, DOPO pulses would oscillate at the state with the minimum loss, which corresponds to the ground state of the Ising Hamiltonian in Equation (1).

In Fig.1B, we present the detailed setup of fs CIM computing architecture. A 1555-nm mode-locked fiber laser with a repetition frequency of 100 MHz and a pulse width of ~100 fs is used as the source, featuring excellent timing stability with a timing jitter < 2 fs. The laser provides an average output power exceeding 100 mW, corresponding to a pulse energy > 1.0 nJ. This source followed by a Second Harmonic Generation (SHG) stage that produces 777.5-nm pulses using a Periodically Poled Lithium Niobate (PPLN) crystal inside the fiber-and-space ring cavity. The crystal has dimensions of 10 mm×1 mm×1 mm, with anti-reflection (AR) coatings on both facets optimized for 1555 nm and 777.5 nm. It has a poling length of 0.5 mm and a poling period ranging from 18.2 μm to 20.9 μm, facilitating efficient quasi-phase-matching for frequency doubling. The DOPO relies on a separate PPLN crystal with dimensions of 10 mm×1 mm×5 mm.

The fiber loop includes two couplers, one for injecting feedback signal pulses and the other coupler for outputting DOPO signals. Laser pulses generated by the fs laser are divided into two paths. One path goes through an Erbium-Doped Fiber Amplifier (EDFA) and an Intensity Modulator (IM) to serve as a feedback injection laser. The other path serves as the local signal for Balanced Homodyne Detection (BHD). Both the injection signal path and the local signal path use fiber stretchers to ensure the alignment between the pump signal and the injection signal in the cavity, and the alignment between the local signal and DOPO output signal. The total number of DOPO pulses in our system is 211, 100 of which are used as signal bits for computation and the rest are used as auxiliary bits to stabilize the system. With the help of Analogue-to-Digital Converter (ADC), signal bits are fed into the Field Programmable Gate Array (FPGA), where the feedback injection strength of each pulse is calculated using the formula $f_i = -r\sum_j J_{ij}\sigma_j$ ($r$ denotes a constant that determines the coupling strength [14]). The measurement and feedback (MFB) scheme are achieved by loading the feedback signal $f_i$ onto the injection laser pulses through a push-pull modulator [13,16,19]. In Fig.1A, a cloud service is set up to host various requests from users of different industries. An Ising model can be created and submitted to the cloud using the Kaiwu SDK [28], and a cloud platform delivers sustained, stable, and task-oriented quantum computing services on CIM machine to users. Fig.1C illustrates the evolution of the in-phase component of each OPO pulse as a function of circulation counts when solving Möbius Ladder graph. In this experiment, we find that the ground state of the Hamiltonian can be found within 20 circulations (<50 microseconds).

# CIM with femto-second pulses

In previously demonstrated CIMs [13,14,16], pico-second (ps) pulses are used for DOPO generation. Here, we found that fs pulses would be more efficient for CIM operations. Compared with CIM with ps pulses, fs pulses result in much higher peak power than that of ps pulses for the same average power. This gives rise to two advantages: (1) more pronounced quantum effects during DOPO generation and (2) lowered pump power during CIM computation. These improvements influence the computation process of the Ising machine, thereby indirectly enhancing the solution quality of combinatorial optimization problems as shown in the next section.

The Hamiltonian of a DOPO in the interaction representation is shown in Equation(2) [29,30]:

$$\hat{H} = \hat{H}_{int} + \hat{H}_{irr} \tag{2}$$

$$\hat{H}_{int} = i\hbar \frac{\kappa}{2} \left( \hat{a}_s^{\dagger 2} \hat{a}_p - \hat{a}_p^{\dagger} \hat{a}_s^2 \right) + i\hbar F \left( \hat{a}_p^{\dagger} - \hat{a}_p \right) \tag{3}$$

$$\hat{H}_{irr} = i\hbar \sqrt{\gamma_s} \left( \hat{a}_s^{\dagger} \hat{B}_s - \hat{a}_s \hat{B}_s^{\dagger} \right) + i\hbar \sqrt{\gamma_p} \left( \hat{a}_p^{\dagger} \hat{B}_p - \hat{a}_p \hat{B}_p^{\dagger} \right) \tag{4}$$

$\hat{H}$ is the Hamiltonian of DOPO, which includes two terms, $\hat{H}_{int}$ and $\hat{H}_{irr}$. $\hat{H}_{int}$ describes the coherent internal dynamics of the DOPO. $\hat{H}_{irr}$ captures the irreversible processes related to energy dissipation and coupling with external reservoirs. $\hat{H}_{int}$ comprises two distinct terms (Equation(3)). The first term represents the non-linear interaction between the signal and the pump. The second term represents the excitation of the pump by an external field denoted $F$. Meanwhile, $\hat{H}_{irr}$ indicates the dissipation processes that affect both the pump and the signal. The term $B_k$, where $k$ can be either $s$ (signal) or $p$ (pump), represents the coupling between these fields and the reservoir field $b_j$. $B_k$ is given by the sum $\sum_j g_{k,j} b_j e^{i(\omega_k - \omega_j)\tau}$. Furthermore, $\hat{a}_s$ and $\hat{a}_s^\dagger$ serve as the creation and annihilation operators for the signal field, respectively. Similarly, $\hat{a}_p$ and $\hat{a}_p^\dagger$ perform the same functions for the pump field. The constant $\kappa$ represents the strength of the non-linear interaction between the signal and pump fields. This term is responsible for the parametric amplification process in DOPO, where the energy from the pump field is transferred to the signal field. The constants $\gamma_s$ and $\gamma_p$ represent the damping rates or dissipation coefficients for the signal and pump fields, respectively. These terms describe the loss of energy from the system due to interactions with the reservoir field ($B_k$). In other words, they quantify how quickly the energy in the signal and pump fields is dissipated into the environment.

As shown in previous work [29,30], the quantum master equation of a DOPO derived from Equation (2) is

$$\frac{\partial \hat{\rho}}{\partial t} = \frac{S}{2}\left[\hat{a}_s^{\dagger 2} - \hat{a}_s^2, \hat{\rho}\right] + \left(\gamma_s \left[\hat{a}_s, \hat{\rho}\hat{a}_s^\dagger\right] + \frac{B}{2}\left[\hat{a}_s^2, \hat{\rho}\hat{a}_s^{\dagger 2}\right] + h.c.\right) \quad (5)$$

where $[A, B] = AB - BA$ represents the commutation relations. $S = \kappa F / \gamma_p$ denotes the squeezing or antisqueezing rate induced by the parametric interaction between the pump and the signal. Compared with ps pulse pumping, fs DOPO results in a larger amplitude $S$ due to the higher conversion rate of the pump energy to the signal [2,29]. Fig.2A shows the Wigner function of DOPO for $S = 1, 1.25, 1.5, 2$. When $S$ increases, the distinction between the spin = −1 and spin = +1 components of DOPO above the threshold becomes more pronounced. This amplifies the impact of quantum noise during CIM calculations.

Another advantage of fs pulses in DOPO is reduced pump threshold power during CIM computation. Under the same average power and repetition frequency, fs pulses possess higher peak power than ps pulses. This gives rise to higher nonlinear conversion efficiency [31]. Fig.2B shows the comparison of SHG efficiency between fs and ps pulse across various pump powers, revealing that the SHG efficiency for fs laser is 2.5 times that of ps laser at the pump power of 120 mW. Therefore, when employing fs pump pulses, the DOPO with higher nonlinear conversion efficiency and shorter pulse duration will operate at lower pump power [13,32]. Fig.2C further examines the output spectra of the OPO at various pump powers. A noticeable spectral broadening can be observed as the pump power increases from 201 to 346 mW. For pump power over 226 mW, the OPO transits from degenerate state to non-degenerate state. Therefore, DOPO operation at low pump power is essential to ensure that CIM operates in the degenerate state. Besides, precise control of pump power is key to the success of CIM calculations. Thanks to the reduced threshold for the fs CIM system, we can better maintain DOPO degeneracy at the lowered pump power.

The stability of CIM is a critical aspect that determines its performance in solving optimization problems. Due to the narrower pulse width with fs pulses, the fs CIM is more sensitive to the length variation in the fiber-and-space ring cavity for DOPOs. Temperature fluctuations cause expansions in optical fibers, which affect the stability of cavity length and coupling between different DOPO pulses. Such instability could result in inferior computational results.

To address this issue, we developed a new cavity stabilization system that integrates temperature control, vibration isolation, and active feedback. To resist temperature fluctuations, we place the fiber in an incubator to isolate it from environmental temperature changes and use an active temperature control system to maintain the temperature inside the incubator. For vibration isolation, acoustic foam is utilized to shield optical components from external vibrations. The effect of this new cavity stabilization system is shown in Fig.3D. For a total testing period of 4 hours, the average 30-minute temperature fluctuation measurements at the optical space cavity and fiber cavity can be reduced to 0.111°C and 0.010°C, while the environment temperature fluctuates around 1.188°C. This is critical for maintaining the stability of fs CIM system and improving the computation accuracy. In addition, active feedback mechanism is deployed to make the system adaptive to short term external noise through a dither-and-lock scheme using auxiliary bits. The auxiliary bits are used to detect the degree of misalignment between DOPO pulses from different circulations at sub-micrometer scale. The misalignment information is then given to an FPGA system that would adjust the cavity length through fiber stretchers according to offset changes induced by external noises. Based on these stabilization measures, our fs CIM system can sustain stable operations for a period of over 8 hours, as discussed in the next section.

**Benchmark on Max-Cut problems**

The Max-Cut problem aims to partition vertices of an arbitrary graph into two sets such that the number of edges between the two sets is maximized. Due to its complexity (NP complete), Max-Cut problem is widely used in many literatures to benchmark the performance of CIMs [12–14,16]. In this work, we tested Möbius Ladder graphs and random graphs of varies sizes on our fs CIM.

1. Möbius Ladder graphs of various sizes

Due to its symmetric nature, Möbius graphs of different sizes all have deterministic optimal solutions, as shown in Fig.4A and Fig.4B. To benchmark the performance of fs CIM, multiple sets of 100 runs were performed for each graph ($V = 20, 40, 60, 80, 100$) and the success rates (which are defined as the probability of achieving the optimal solution) are shown in Fig.4C. The optimal solution can be found consistently for each graph, as the success rate remains positive for all graphs. In particular, for the number of vertices less or equal to 80, the average success rate remains above 60%. The error bars denote the variability in the success rate, typically ranging from 0% to 10%.

As shown in Table 1, compared with previously reported success rates [13] and the simulated annealing (SA) algorithm, our fs CIM achieves the highest average success rate (55% vs 20%, 2%) for solving the 100-vertex Möbius Ladder graph, while consuming the shortest computation time (464 μs vs 480 μs, 120 ms per run). The details of the experiment can be found in supplementary material.

2. Random graphs with different edge densities

Compared with Möbius Ladder graphs, random graphs are generally difficult to compute, and the optimal solution is unknown for most of the problems. In real-world situations, the edge density distribution of optimization problems is relatively broad. To better approximate the edge density distribution of real-world problems, we sampled random graphs in a uniform manner with edge densities d ($d = 2|E|/V(V-1)$) ranging from 1% to 99% (the actual edge densities of the generated problems might deviate from target values due to the inherent randomness in the sampling program). As shown in Fig.5A, a series of 100-vertex random graphs with edge densities of 1.7%, 19.6%, 39.8%, 60.5%, 78.5%, and 98.9% are experimented on our fs CIM to show the capability of the system to solve complex issues (the algorithm for generating random graphs can be found in GitHub [33]). The Ground State (GS) solutions were computed via Biq Mac Solver [34] and the max-cut values at GS are shown in Fig.5B. To better access our fs CIM's performance, we further define 98% GS and 95% GS as cut values that are 98% and 95% of GS cut values or higher. Fig.5C shows the success rate of fs CIM achieving GS, 98% GS, 95% GS for various random graphs.

For GS computation, the success rate gradually decreases as the edge density increases. The success rate is over 80% for sparse graph (edge density = 1.7%), and remains above 40%, as the edge density goes to ~40%. As edge density increases, the number of possible configurations grows exponentially, making it harder for CIMs to navigate the solution space and guarantee an exact optimal cut. This combinatorial explosion reduces the success rate monotonically. As a result, the success rate decreases as the edge density increases from 10% to 80%.

For approximate solutions like 98% GS and 95% GS, similar trends can be observed, except for edge density over 80%. When edge density is over 80%, the success rates for 98% GS and 95% GS increase gradually, unlike the monotonic decrease for GS computation. In sparse graphs (low density <40%), optimal cuts could rely on exploiting specific structures, which may allow for significant improvements over random partitions. However, these structures become less pronounced as density increases. At medium densities (~40-80%), the solution space would increase dramatically and the energy landscape for optimization becomes rugged with more local minima, causing CIMs to struggle even for near-optimal solutions. The gap between average and optimal cut values is large, making approximations challenging. However, in very dense graphs (density >80%), the structure becomes uniform, resembling a complete graph. Here, the max-cut value stabilizes near $n^2/4$ (for $n$ vertices), and even random partitions achieve ~50% of edges cut. This uniformity reduces variability in cut values. Therefore, the optimal cut in dense graphs is only marginally better than typical cuts, so achieving a 98% approximation becomes easier. Compared to GS, success rate for 98%GS and 95%GS are higher, and can stay above 70% and 80% respectively. These results are also higher than the previously reported data [13], demonstrating the improved performance for our fs CIM with stabilization measures. For solving random graphs at 60% edge density, the best success rates in prior work [13] reached only ~10% (with worst-case at 0%), whereas our fs CIM consistently achieves success rates above 10% even in its worst cases.

3. Long-term stability of the CIM

The stability of CIM is crucial for its deployment for large scale computing applications. To evaluate system's performance over an extended period, a random graph with 100 vertices and 700 edges is employed as a testing problem (shown in Fig.3A). This testing problem is being calculated on our fs CIM continuously for 8 hours. The results are shown in Fig.3B, out of 18000 calculations conducted during the 8-hour period, there is a 50.58% probability of finding the optimal solution with the max-cut value of 475 and a 98.58% probability of finding a solution that has a cut value not less than 471 (99% of the optimal cut value). In only a few instances (7 out of 18000), the system returned a solution whose cut value is less than 90% of the optimal cut value (428). Toward the end of the 8-hour period, the results did not show any degradation in solution quality. This suggests that our system is likely to maintain stable operation over the test period of 8 hours.

**Solving QUBO problems for practical applications**

Quadratic Unconstrained Binary Optimization (QUBO) is used in a variety of applications, such as scheduling, routing, and machine learning [15,21,35–38]. QUBO problems can also be effectively transformed into the Ising models (using method described in Supplementary Materials), which then can be solved using our fs CIM. Here we demonstrate two application areas exploiting the performance benefits of fs CIM.

1. Molecular docking

As a fundamental and essential technology, molecular docking is widely used in drug discovery and biological research and aims to identify an optimal binding pose based on a minimum global energy in ligand-protein complexes (Fig.6A) [36,39]. Docking involves two main processes: sampling and scoring, where the sampling process is a classic NP-hard problem, and traditional algorithms are unable to calculate accurate results in a vast search space. Our previous studies encoded the sampling process into a QUBO problem and demonstrated its performance on benchmark datasets through simulators [35]. In this paper, we used our fs CIM to solve the molecular docking problems. As shown in Fig.6C, the data are acquired from the PDB database [37] and are preprocessed to protein and ligand files using prepare receptor and prepare ligand tools in AutoDockFR 1.0 [37,38]. Then, we use Kaiwu SDK [28] to construct QUBO models first and converted them into the Ising Hamiltonian according to Equation(6). In this model, we transform the molecular docking into a problem of matching individual atoms in the molecule to discretized protein lattice points. $x_{ij}$ in this QUBO model is defined as the binary decision variable to indicate whether the atom $a_i$ and the grid point $g_j$ are matched. $w_{a_i,g_j}$ denotes the energy coefficient for this matching case, and $K_{dist}$ and $K_{memo}$ represent the penalty coefficient. $u_{ijkl}$ and $v_{ijkl}$ are two hard constraints that limit the distance between an atom and a lattice point match to be legal and limit the existence of at most one match for a lattice point and an atom, respectively. More details can be found in [35].

$$H = \sum_{i=1}^{n}\sum_{j=1}^{N} w_{a_i,g_j} x_{ij}^2 + K_{dist} \sum_{i=1}^{n}\sum_{j=1}^{N}\sum_{k=1}^{n}\sum_{l=j+1}^{N} u_{ijkl} x_{ij} x_{kl} + K_{mono} \sum_{i=1}^{n}\sum_{j=1}^{N}\sum_{k=1}^{n}\sum_{l=j+1}^{N} v_{ijkl} x_{ij} x_{kl} \quad (6)$$

The quality of poses sampled by CIM is then evaluated by the Root Mean Square Deviation (RMSD) between sampling poses and crystal structures. We choose the minimum RMSD (mRMSD) in one computation as the CIM solution. In molecular docking, it is generally accepted that an mRMSD less than 2Å is an acceptable docking result.

Three systems, named 1N2J, 1LRH, and 1JD0, are constructed with QUBO models and solved via our fs CIM. Together, these three systems cover a wide spectrum of biological functions and organism types (supplementary material), enabling multiple validation of the proposed computational approach. Fig.6B illustrates the evolution of the Hamiltonian in the solution process, demonstrating that the CIM machine can rapidly acquire docking results in 300 micro-seconds. Fig.6C and Table 2 show calculated mRMSD results using fs CIM. mRMSD values for 1N2J, 1LRH, and 1JD0 are 0.8Å, 1.4Å, and 0.6Å respectively. All these are high-quality (mRMSD < 2Å) docking poses, which leave very small differences between the sampling poses and crystal structures. This demonstrates that our fs CIM can effectively solve the molecular docking problem and our quantum docking method can obtain good sampling poses in a sub-ms scheme.

This demonstrates that our fs CIM can effectively solve the molecular docking problem and our quantum docking method can obtain good sampling poses in a sub-ms scheme. Our platform integrates advanced computational algorithms with optimized workflows, resulting in improved computational efficiency without compromising accuracy. This balance enables rapid screening and analysis of molecular systems at scales that are often challenging for other platforms. Moreover, the platform demonstrates enhanced flexibility and scalability, allowing it to accommodate a diverse range of molecular targets and binding scenarios. Through multiple benchmarks on representative test systems, our platform consistently achieves comparable or superior predictive performance, as evidenced by standard metrics such as RMSD and binding affinity correlations.

2. Credit scoring and classification

Credit scoring and classification plays an important role in computational finance [40]. The primary objective is to predict the credit worthiness of new applicants, leveraging data from historical credit applications [41]. This process encompasses several steps such as feature selection, data cleaning, and classification [42]. Among these, feature selection is a crucial step that aims to reduce the number of variables for a classification model (which will be trained using algorithms like XGBoost [43]), and further improving scoring accuracy. In this work, we utilize our fs CIM to select proper features for credit scoring. The QUBO expression for feature selection can be expressed as follows:

$$f(x) = -\left( \alpha \sum_{j=1}^{n} x_j |\rho V_j| - (1-\alpha) \sum_{j=1}^{n} \sum_{k=1, k \neq j}^{n} x_j x_k |\rho_{jk}| \right) \quad (7)$$

$x_j$ is a binary decision variable that indicates whether the $j$-th feature is selected ($x_j = 1$) or not ($x_j = 0$). It directly influences whether a particular feature will be included in subsequent analysis or model training. $\alpha$ is a weighting factor that balances the trade-off between individual features and the inter-feature correlations. It ranges between 0 and 1. A higher value of $\alpha$ emphasizes the importance of individual features' contributions to the label category, while a lower value gives more weight to minimizing the correlations among different features, promoting feature independence. $n$ represents the total number of features considered for selection. The summation indices $j$ and $k$ range from 1 to $n$, ensuring that all features are evaluated within the QUBO framework. $\rho V_j$ captures the influence or relevance of the $j$-th feature in the label category (e.g., credit reliability). It could be a measure such as the feature's correlation with the target variable, mutual information, or any other metric that quantifies the feature's importance. The absolute value $|\rho V_j|$ ensures that the impact is measured in absolute terms. $\rho_{jk}$ represents the correlation between the $j$-th and $k$-th features. It measures the degree of linear dependency or similarity between two features. The absolute value $|\rho_{jk}|$ is used to ensure that the correlation strength is considered regardless of its direction (positive or negative). Minimizing this term helps reduce redundancy and promote feature independence [44]. This formulation guides the CIM in selecting an optimal subset of features, which can then be used effectively in training models for credit scoring and classification tasks.

We construct the QUBO model based on Kaggle dataset [45] using the aforementioned approach. The QUBO formulation was then converted into an Ising problem using Kaiwu SDK, yielding the $J_{ij}$ matrix for calculation.

To demonstrate the advantage of fs CIM over classical optimizers, we compared the performance of XGBoost with feature selection (CIM + XGBoost model) to that of XGBoost without CIM assisted feature selection (XGBoost only model). Fig.7A illustrates the scoring process that aims to separate the overdue cases from normal repayments. For a given scoring model, distributions of normal and overdue repayments can be plotted (Fig.7A) and the non-overlapping areas between these two distributions indicate the effectiveness of a scoring method.

Kolmogorov-Smirnov (KS) value is used to quantifies the effectiveness for different scoring methods by measuring the maximum difference between the cumulative distributions of normal and overdue repayments (as illustrated in Fig.7B) [46]. A higher KS value indicates a greater discriminatory power of the model [47].

In Fig.7C, experimental results of CIM + XGBoost model are shown and compared to the traditional method of using XGBoost only model. For our CIM + XGBoost model, weighting factor α is used to fine tune the model and the resulting KS values are shown in Fig.7C. As α increases, the KS statistic of the CIM + XGBoost mothed rises almost monotonically. CIM + XGBoost approach achieves the maximum KS value of 0.3817 at $\alpha = 0.9975$, outperforming XGBoost only model, which only attains a KS value of 0.3767. By facilitating feature selection, fs CIM enables the removal of invalid or potentially detrimental features, thereby enhancing the generalizability of the scoring model.

To sum up, our platform leverages a quantum feature selection approach that fundamentally differs from classical methods (e.g., embedded, wrapper, or filter techniques). By formulating feature selection as a QUBO (Quadratic Unconstrained Binary Optimization) problem, the method inherently captures complex interdependencies among features. This allows the platform to evaluate high-order feature correlations in parallel, significantly accelerating the identification of the most informative and non-redundant feature subsets. As a result, the quantum-enhanced workflow not only improves selection efficiency but also enhances model interpretability and predictive performance compared to conventional approaches.

**Discussion**

1. Femto- vs pico-second laser

    The narrower width of fs pulses results in higher peak power density, which may enhance the nonlinear conversion efficiency during down-conversion process. As a light source with high power, it also exhibits more pronounced quantum effects. This improvement influences the the process of computation of the Ising machine, thereby indirectly enhancing the solution quality of combinatorial optimization problems. The fs pulsed laser used in our experiment has a pulse width of <300 fs, approximately 1/50 of the 15 ps pulse width reported in previous studies [13] . Phase locking of the pulses is required during system operation; consequently, such a narrow pulse width is highly sensitive to DOPO cavity length variations and environmental noise in the time domain. To address this, we developed a novel cavity stabilization system to shield it from environmental temperature fluctuations, adopted an active temperature control system to maintain the incubator's internal temperature, and used vibration-damping foam for vibration isolation to protect optical components from external vibrations.

2. Scalability and limitations

    To scale the system to a larger size, two approaches can be adopted: (1) increasing the repetition rate of the fs laser; (2) extending the fiber length. However, the

repetition rate of fs laser cannot be increased arbitrarily and longer fiber length incurs higher cavity loss and makes system more sensitive to environment noises. While many key components have been integrated into chips to improve system's stability[48], kilometer scale on-chip fiber fabrication have not yet been achieved.

3. Conclusion

CIMs provide a promising platform for solving QUBO and Ising problems across various domains, offering advantages in speed, efficiency, and the ability to address large-scale optimization tasks. In summary, we achieved 100-spin CIM with optimized performance by unitizing fs DOPOs. Furthermore, our fs CIM system can perform stable calculations for over 8 hours and opens an new path to improve the performance of CIMs. Through the test runs of problems from molecular docking and credit scoring, fs CIM further demonstrates its potential in many optimization areas. This encourages us to consider the large-scale deployment of fs CIM as a potential alternative technology to tackle complex optimization issues in the future.

**Acknowledgments**

**Funding:**

National Natural Science Foundation of China grant 62371050

National Natural Science Foundation of China grant 62131002

Fundamental Research Funds for Central Universities (BNU)

**Author contributions:**

Conceptualization: HW, CW, KW

Methodology: HW, AC, PG, BJ, CC, JW

Investigation: AC, PG, BJ, YL, HS, SC, CC, CJ, JW, YM

Visualization: PG, SC

Supervision: HW, CW, KW

Writing—original draft: PG, HS, YL, SC, CJ

Writing—review & editing: HW, KW

**Competing interests:** Authors declare that they have no competing interests.

**Data and materials availability:** All data are available in the main text or the supplementary materials. The code for molecular docking and credit scoring can be accessed at https://github.com/qboson/100-vertex-random-graphs.git.

**Fig. 1. The computing architecture of fs CIM.** (**A**) Job execution flow chart of fs CIM computing services. The complete process involves the user submitting computing jobs, such as molecular docking, financial analysis, neural networks, and material design, to a cloud server. The cloud server then converts these jobs into an Ising model matrix and transmits it to the local computer of the CIM via Kaiwu SDK (25). The FPGA is connected to the local computer for calculation using the obtained matrix. (**B**) Sketch of the overall architecture of fs CIM. Device description. PPLN: periodically poled lithium niobate, IM: intensity modulator, EDFA: erbium-doped fiber amplifier, SHG: second harmonic generation, PZT: piezoelectric transducer, AOM: acoustic optical modulator, BHD: balanced homodyne detection. (**C**) The evolution of the in-phase components of the $N = 100$ OPO pulses as a function of the circulation times in solving the Möbius Ladder graph.

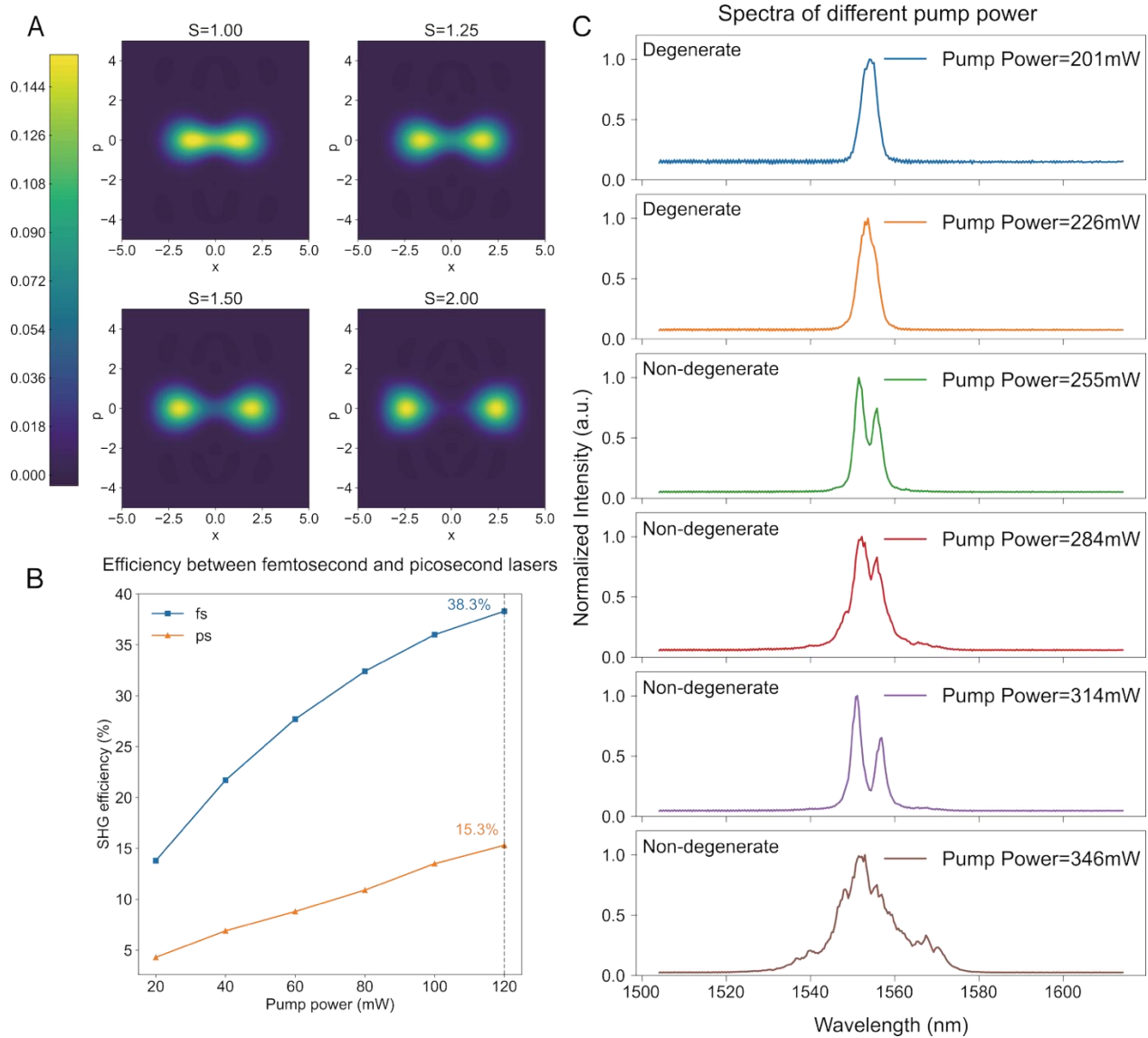

**Fig. 2. The advantages of fs CIM.** (**A**) Wigner functions of DOPOs above the threshold. The horizontal axis represents the position variable, while the vertical axis corresponds to the momentum variable in phase space. These axes define a two-dimensional plane where the Wigner function provides a quasi-probability distribution, capturing both quantum and classical features of the state. S is the second-order non-linear coupling constant between the signal and pump. When S increases, the differentiation between the spin = −1 and spin = +1 components of DOPO becomes more pronounced. (**B**) The measured SHG efficiency by pumping using fs lasers and ps lasers. The pulse width for the fs laser is <300 fs, whereas the pulse width for the ps laser is ~3 ps. The repetition frequency is kept at 100 MHz for both measurements. At a pump power of 120 mW, the SHG efficiency reaches 38.3% for the fs laser and 15.3% for the ps laser, showing a 2.5X increase in efficiency. (**C**) Output spectra of the DOPO/OPO at different pump powers. When the pump power increases from 201 mW to 226 mW, the DOPO spectrum begins to widen. Once the pump power exceeds 226 mW, and the state of OPOs shifts from degeneracy to non-degeneracy.

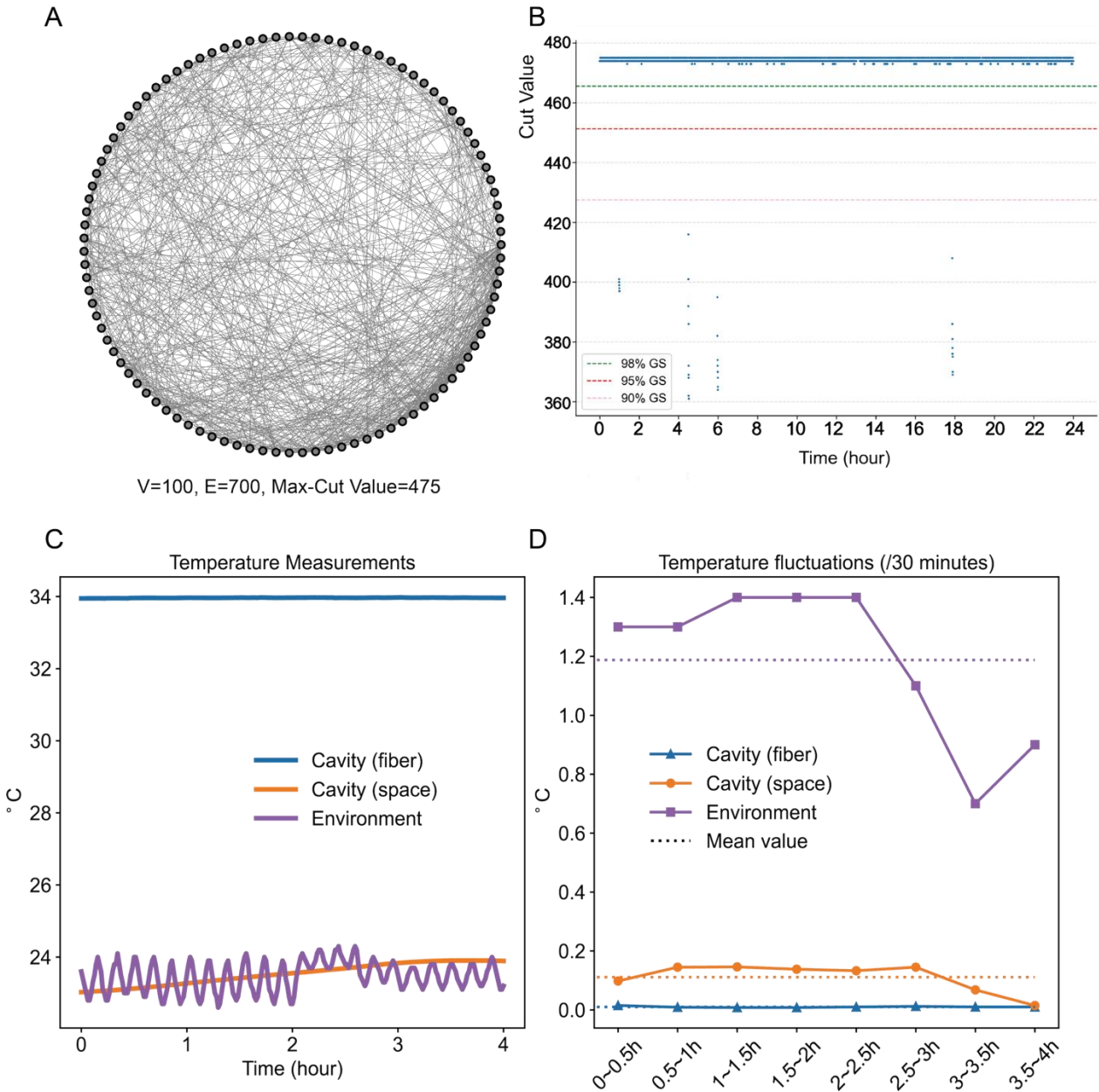

**Fig. 3. Demonstration of stable CIM.** (**A**) The test problem with its GS cut value. (**B**) Long-term computing performance of fs CIM. The same problem is being calculated on the fs CIM and resulting cut values are recorded for over 8 hours.100-vertex random graphs with different edge densities (=1.7%, 19.6%, 39.8%, 60.5%, 78.5%, 98.9%). (**C**) Thermal measurement of the spatial optical path and the surrounding fiber before and after temperature control. The environmental thermal fluctuation can be reduced with the help of the temperature control system. (**D**) Temperature fluctuations (/30 minutes) of the spatial optical path and the surrounding fiber before and after temperature control. Detailed numerical data are provided in Table. S1 of the Supplementary Materials.

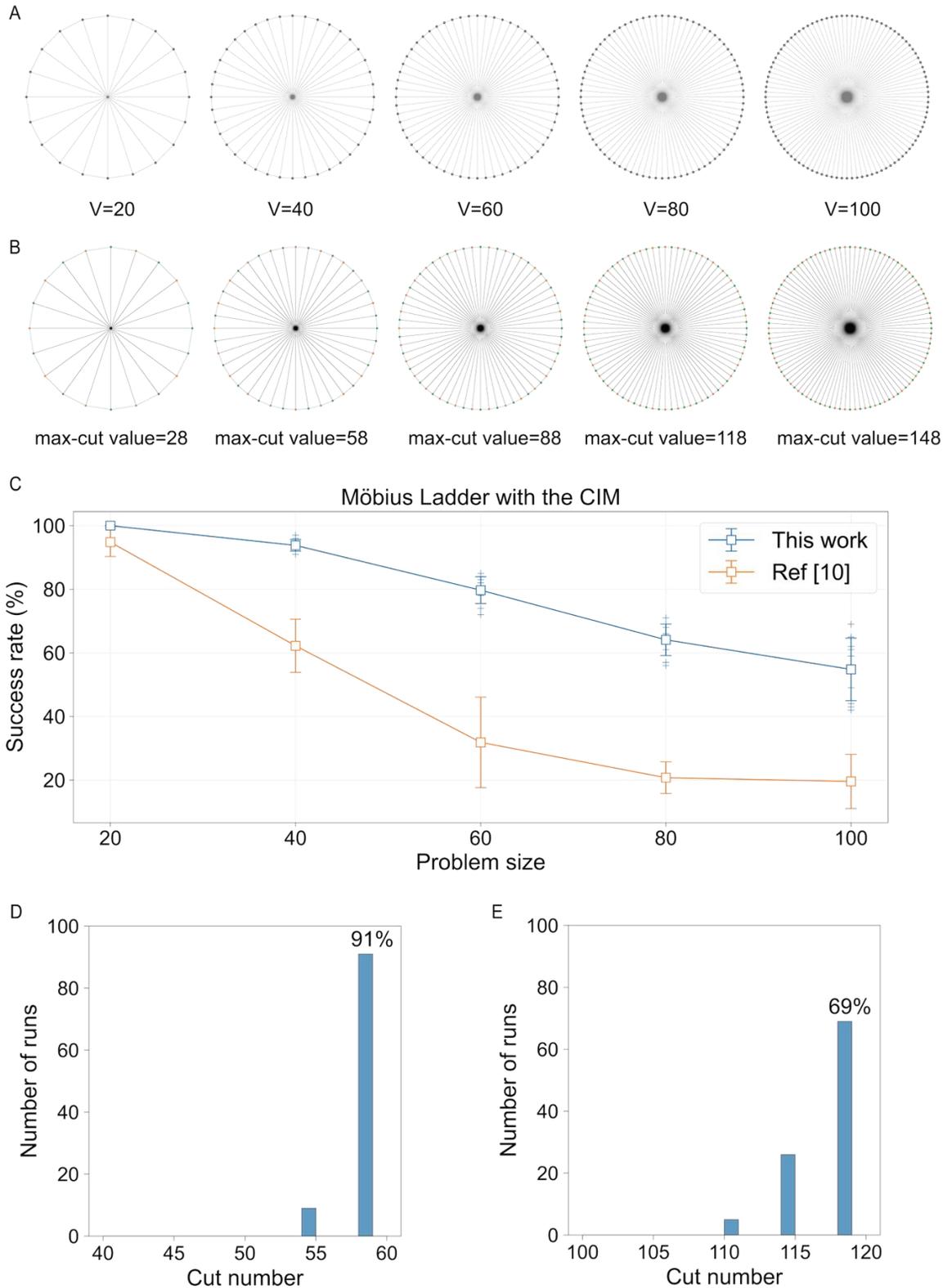

**Fig. 4. Benchmark Analysis of Möbius Ladder Graphs.** **(A)** Möbius Ladder graphs with different numbers of vertices (V = 20, 40, 60, 80, 100). **(B)** The ground state solutions and resulting max-cut values for the Möbius Ladder graphs in (A). **(C)** Max-cut results for various Möbius Ladder graphs. Measured probability of achieving the ground state in a single run, as a function of number of vertices. Multiple 100-run batches were performed for each graph to obtain the standard deviations, which are shown as error bars. **(D)** Histogram of obtained solutions in 100 runs for Möbius Ladder graphs with 40 vertices. **(E)** Histogram of obtained solutions in 100 runs for Möbius Ladder graph with 80 vertices.

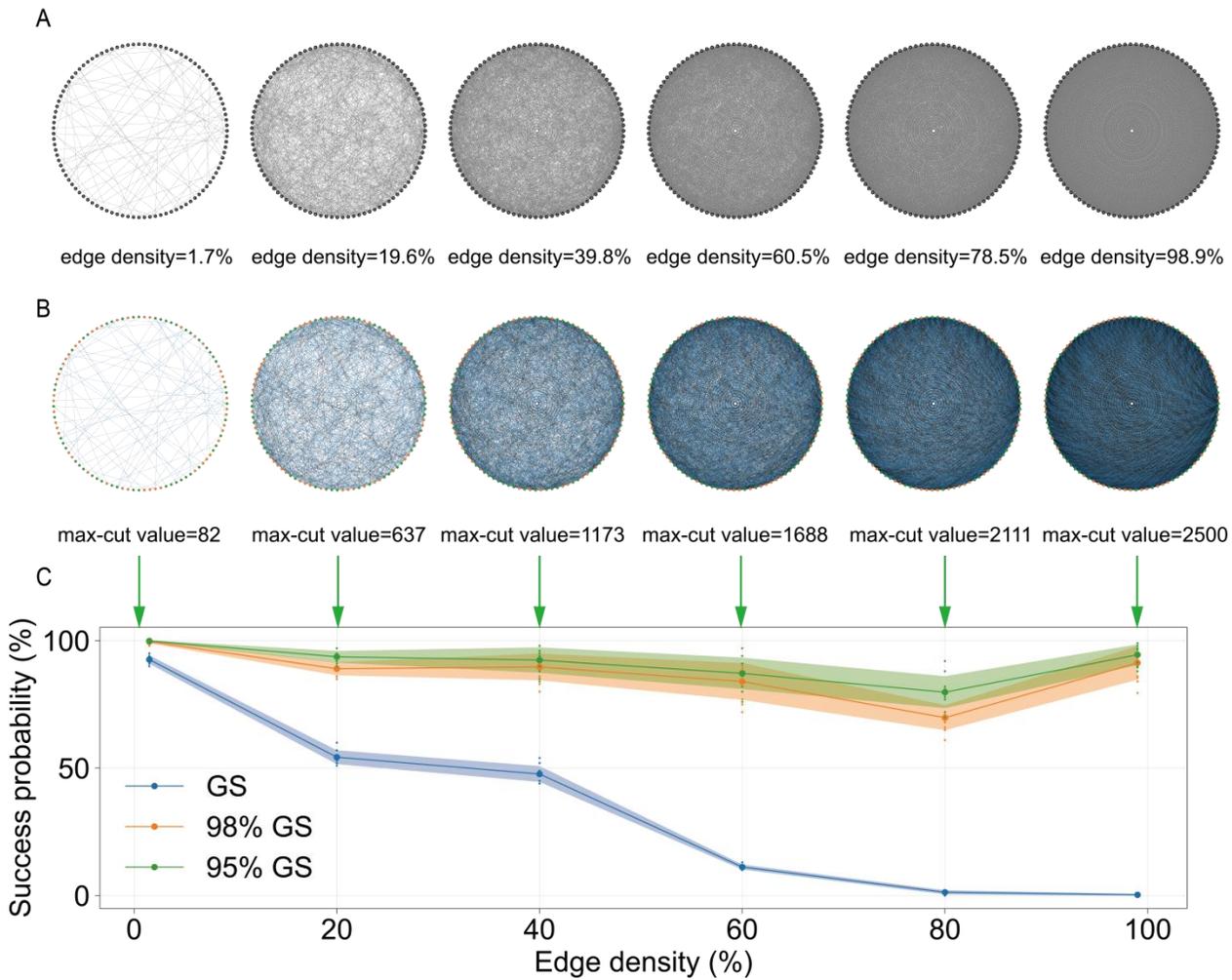

**Fig. 5. Benchmark Analysis of Random Graphs.** (**A**) 100-vertex random graphs with different edge densities (=1.7%, 19.6%, 39.8%, 60.5%, 78.5%, 98.9%). (**B**) Ground state solutions for graphs in Fig.5A and the resulting max-cut values. (**C**) Results with various-density random graphs. Observed probability of obtaining a solution whose cut size is at least x% of the max-cut at ground state, as a function of edge density. The running graphs are random graphs with 100 vertices and various densities. Error bars indicate the standard deviations in multiple 100 runs.

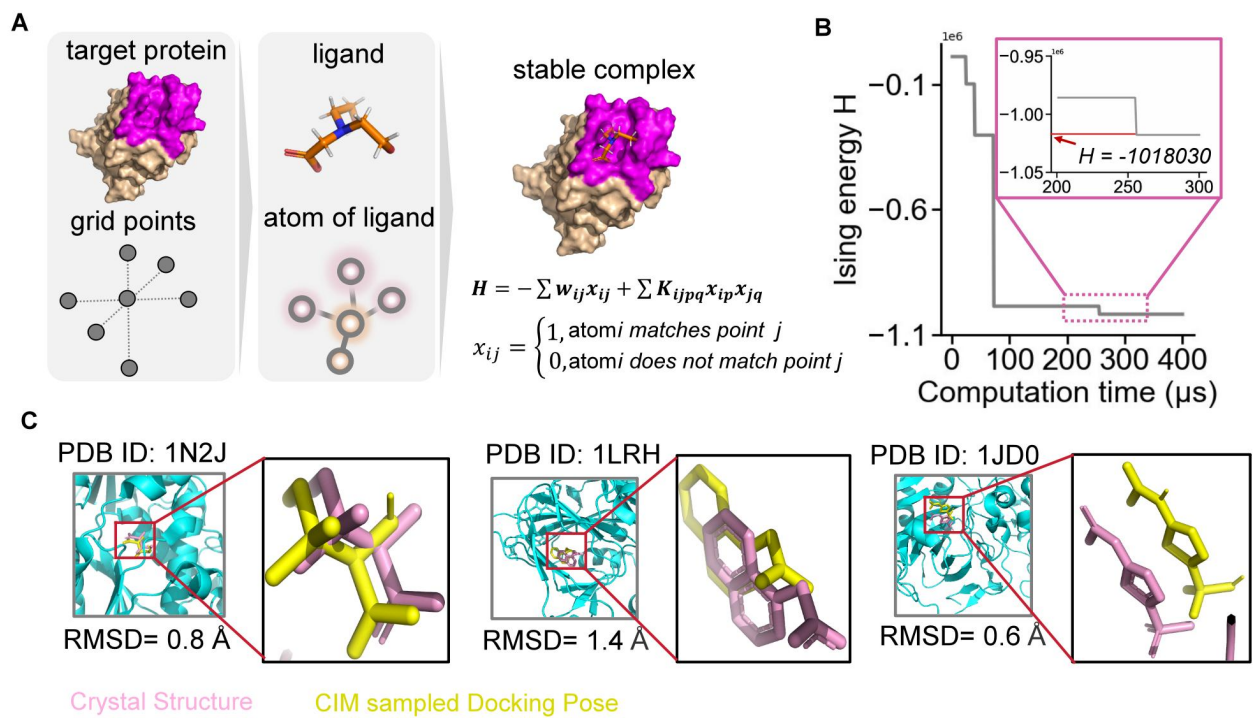

**Fig. 6. Molecular docking by CIM.** (**A**) Encoding the sampling process of molecular docking into a QUBO model. (**B**) Ising energy evolution of the CIM single solution. (**C**) 1N2J, 1LRH, and 1JD0 of the sampled docking poses (yellow) compared with the poses in the crystal structures (pink). The mRMSD values for these three molecules are 0.8Å, 1.4Å, and 0.6Å, respectively, all of which are below 2Å, indicating the effectiveness of CIM in molecular docking calculations.

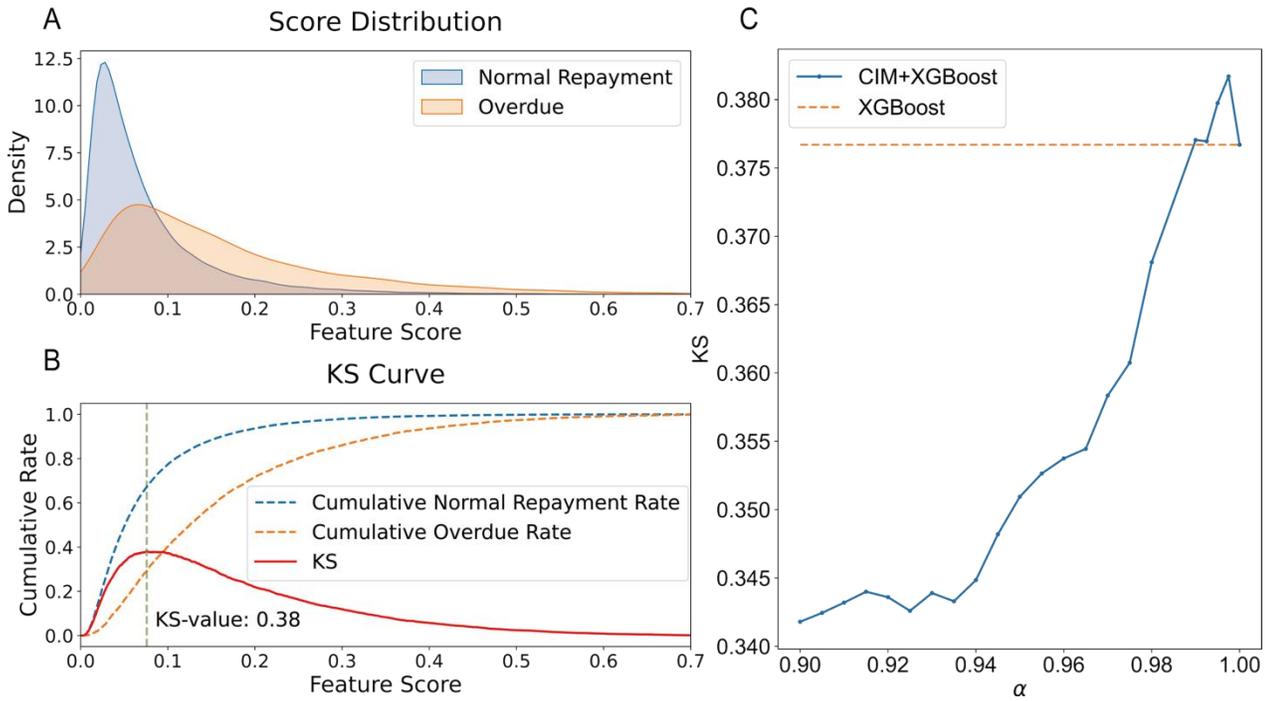

**Fig. 7. Credit scoring by CIM.** (**A**) The score distributions for normal and overdue repayments. The non-overlapping area between the two distributions indicates the effectiveness of the feature selection process. (**B**) Definition of KS (Kolmogorov-Smirnov) statistic, which measures the maximum difference between the cumulative distributions of different labels. (**C**) As the alpha parameter increases, the KS statistic of the quantum feature selection CIM + XGBoost model rises. When alpha exceeds 0.99, the quantum feature selection CIM + XGBoost model outperforms the XGBoost model without feature selection.

| Instances | Success rate of this work (464 μs) | Success rate of CIM [10] (480 μs) | Success rate of SA (~120 ms) |
|---|---|---|---|
| Möbius Ladder graph with 20 vertices | 100% | 95% | 100% |
| Möbius Ladder graph with 40 vertices | 94% | 62% | 69% |
| Möbius Ladder graph with 60 vertices | 80% | 32% | 23% |
| Möbius Ladder graph with 80 vertices | 64% | 21% | 8% |
| Möbius Ladder graph with 100 vertices | 55% | 20% | 2% |

**Table 1. Success rate of Möbius Ladder graphs of various sizes for this CIM (464 μs), the CIM 10 (480 μs), and SA (~120 ms).**

| PDB ID | mRMSD (Å) | If mRMSD < 2Å |
|--------|-----------|---------------|
| 1N2J   | 0.8       | Yes           |
| 1LRH   | 1.4       | Yes           |
| 1JD0   | 0.6       | Yes           |

**Table 2. mRMSD of three molecules calculated using fs CIM.** PDB ID is identity of the molecule in PDB database. All three mRMSD values are below 2Å, indicating that CIM is effective in solving molecular docking problem.